\documentclass[twocolumn,english,aps,pra,showpacs,superscriptaddress]{revtex4-1}
\usepackage[latin9]{inputenc}
\usepackage{mathrsfs}
\usepackage{amsmath}
\usepackage{amssymb}
\usepackage{graphicx}
\usepackage{esint}

\makeatletter

\usepackage{babel}

\usepackage{babel}

\usepackage{babel}

\makeatother

\usepackage{babel}
\begin{document}

\title{Full-polaron master equation approach to dynamical steady states
of a driven two-level system beyond the weak system-environment coupling}

\author{Chien-Chang Chen}

\affiliation{Department of Physics and Center for Theoretical Physics, National
Taiwan University, Taipei 10617, Taiwan}

\author{Thomas M. Stace}

\affiliation{ARC Centre for Engineered Quantum Systems, School of Mathematics
and Physics, The University of Queensland, Brisbane, Queensland 4072,
Australia}

\author{Hsi-Sheng Goan}

\email{Email: goan@phys.ntu.edu.tw}

\selectlanguage{english}%

\affiliation{Department of Physics and Center for Theoretical Physics, National
Taiwan University, Taipei 10617, Taiwan}

\affiliation{Center for Quantum Science and Engineering, National Taiwan University,
Taipei 10617, Taiwan}

\date{\today}
\begin{abstract}
We apply a full-polaron master equation and a weak-coupling non-Markovian
master equation to describe the steady-state time-averaged properties
of a driven two-level system, an electron coherently tunneling between
double quantum dots (DQDs), interacting with a bosonic phonon bath.
Comparing the results obtained using these two master equations with
those from a recent DQD experiment and its corresponding weak-coupling
theoretical method, we find that the original parameter set used in
the experiment and theoretical method is not in the weak-coupling
parameter regime. By using the full-polaron master equation with a
slight adjustment on only the value of the interdot separation in
the original experimental parameter set, we find that a reasonable
fit to the experimentally measured time-averaged steady-state population
data can be achieved. The adjusted interdot separation is within the
possible values allowed by the geometry of the surface gates that
define the DQD in the experiment. Our full-polaron equation approach
does not require the special renormalization scheme employed in their
weak-coupling theoretical method, and can still describe the experimental
results of driving-induced phonon-enhanced steplike shoulder behaviors
in the experiment. This demonstrates that the full-polaron master
equation approach is a correct and efficient tool to describe the
steady-state properties of a driven spin-boson model in the case of
strong system-environment coupling. 
\end{abstract}
\maketitle

\section{Introduction}

Dynamics of driven open quantum systems is of broad interest and great
importance for many different fields and disciplines ranging from
quantum information processing to biological physics. For a standard
spin-boson model with a two-level energy splitting $\hbar W$ and
a environment-induced decay rate $\Gamma$ proportional to the system-bath
coupling strength $P$, the master equation approach \cite{Carmichael}
via weak-coupling perturbation with respect to $P$ is one of the
most often adopted approaches to treat open quantum systems. This
approach is valid in the weak coupling regime, i.e., small $P$ such
that $W\gg\Gamma$. While for a spin-boson model sinusoidally driven
at frequency $\omega_{0}$, even though $W\gg\Gamma$, this perturbation
process may fail under the weak driving and near or on resonance conditions
\cite{A_general_RWA_VP_master_equation} where $\max(\Omega_{0},|\delta|)\ll\Gamma$,step-like
with $\Omega_{0}$ being the driving amplitude and $\left|\delta\right|=\left|W-\omega_{0}\right|$
being the detuning.

Quantum dots are promising physical systems for coherence control
experiments due to their great controllability and tunability \cite{Ref8_PRL111_Petta2004,Ref12_PRL111_Hanson2007}.
In this paper, we investigate a driven double quantum-dot (DQD) system
in a recent experiment \cite{step-like-experiment2014}. When the
DQD is driven such that the driving amplitude $\Omega_{0}$ is comparable
to the energy scale of system-environmental coupling, a steplike shoulder
may appear on the blue-detuned off-resonant side $\left(W<\omega_{0}\right)$
of the resonance peak. The asymmetry of the line shape of the sidebands
indicates the interaction with the environment as excitation channels
are opened when $\omega_{0}>W$ \cite{StaceDohertyBarrett2005,Stace_steady-state_PRL111_2013,step-like-experiment2014}.
The experiment in Ref. \cite{step-like-experiment2014} showed this
clear asymmetry and steplike shoulder feature on the blue side of
the sideband and theoretical work based on a weak system-environment
coupling theory \cite{Stace_steady-state_PRL111_2013} was put forward
to explain the observed feature in Ref. \cite{step-like-experiment2014}.

In this paper, we use both a weak system-environment coupling treatment
and a full-polaron transformation approach serving as a strong system-environment
coupling treatment to study the driven DQD system interacting with
a phonon bath described in Ref. \cite{step-like-experiment2014}.
We find that the parameters of the driven DQD system used in Ref.
\cite{step-like-experiment2014} are beyond the weak system-environment
coupling regime. However, the theoretical method \cite{Stace_steady-state_PRL111_2013}
employed in Ref. \cite{step-like-experiment2014} to explain the experiment
involves only second-order perturbation theory in the system-bath
interaction. Therefore, using the full-polaron method valid for strong
system-environment coupling, we find that to fit the experimental
data, only one critical parameter, the interdot separation, adopted
in Ref. \cite{step-like-experiment2014} for their theory, is required
to be adjusted by a small magnitude. The adjusted interdot separation
is still consistent with the distance and geometry of the surface
gates that confine and define the DQD system as shown in Ref. \cite{step-like-experiment2014}.

The paper is organized as follows. We describe in Sec.~\ref{sec:driven_DQD_model}
the model Hamiltonian for the driven DQD system, and then present
the derivations of our weak-coupling and full-polaron master equations.
Numerical results are presented in Sec.~\ref{sec:Results}, in which
we compare our two master-equation approaches with the experiment
data \cite{step-like-experiment2014}. A comparison between the approach
in Refs.~\cite{Stace_steady-state_PRL111_2013,step-like-experiment2014}
and our weak-coupling and full-polaron master-equation approaches
is presented in Sec.~\ref{sec:Discussion}. Then a short conclusion
is given in Sec.~\ref{sec:Conclusion}. Finally, we discuss the validity
of the parameter set of the driven DQD system used in Ref.~\cite{step-like-experiment2014}
in Appendix \ref{sec:appendixA_d_20_fit} and the problem of positivity
violations in second-order master equations in Appendix \ref{sec:appendix_positivity}.

\section{Model Hamiltonian and non-Markovian master equation\label{sec:driven_DQD_model}}

The system considered in the experiment \cite{step-like-experiment2014}
is an electron in a DQD system driven by a microwave at frequency
$\omega_{0}$ and amplitude $\Omega_{0}$. There is an energy bias
$\epsilon$ between localized left $\left|l\right\rangle $ and localized
right $\left|r\right\rangle $ states of the DQD, and the electron
tunnels coherently between the DQD with interdot tunneling rate $\Delta$.
Furthermore, the DQD system is coupled to its surrounding bosonic
bath and the total Hamiltonian describing the whole system \cite{Stace_steady-state_PRL111_2013}
is $\left(\hbar=1\right)$: 
\begin{eqnarray}
H_{T}(t) & = & H_{s}(t)+H_{b}+H_{sb},\label{H_T(t)}\\
H_{s}(t) & = & -\frac{\epsilon}{2}\sigma_{z}-\frac{\Delta\left(t\right)}{2}\sigma_{x},\\
H_{b} & = & \sum_{k}\omega_{k}b_{k}^{\dagger}b_{k},\\
H_{sb} & = & \sigma_{z}\sum_{k}g_{k}\left(b_{k}^{\dagger}+b_{k}\right),\label{eq:H_sb}
\end{eqnarray}
where $\sigma_{z}\equiv\left|l\right\rangle \left\langle l\right|-\left|r\right\rangle \left\langle r\right|$,
$\sigma_{x}\equiv\left|l\right\rangle \left\langle r\right|+\left|r\right\rangle \left\langle l\right|$,
and 
\begin{equation}
\Delta\left(t\right)=\Delta-2\Omega_{0}\cos\left(\omega_{0}t\right).\label{eq:Delta(t)}
\end{equation}
Here, the DQD is treated as a two-level system (qubit) with Hamiltonian
$H_{s}$, the Hamiltonian for the bosonic bath is $H_{b}$ with frequency
$\omega_{k}$ and creation (annihilation) operator $b_{k}^{\dagger}\left(b_{k}\right)$
of the bath modes, and $H_{sb}$ is the system-bath interaction Hamiltonian
with strength $g_{k}$ coupling to the respective bath mode $k$.
Next, we will describe how our weak-coupling and the polaron master
equations are obtained and then use them to compare with the experimental
data.

\subsection{Weak-coupling non-Markovian master equation}

\label{sec:weak-coupling}

First, we introduce a master equation valid to second order in system-bath
coupling strength and at the same time valid for a strong driving
field. We will adopt the weak-coupling time-nonlocal (time-convolution)
non-Markovian master equation \cite{Breuer02,time-local_vs_non-loca_2nd_ME_initial}
to describe the time evolution of the reduced density matrix of the
system 
\begin{equation}
\rho_{s}\left(t\right)=\textrm{tr}_{b}\left[\rho_{T}\left(t\right)\right]\label{rhos(t)}
\end{equation}
to compare directly with that in Ref. \cite{Stace_steady-state_PRL111_2013}
as a time-nonlocal non-Markovian master equation was used there. Because
we focus on the comparison of the steady-state population with the
experiment \cite{step-like-experiment2014}, and in our case the dynamical
steady-state result is independent of any reasonable choices of initial
states, the initial total density operator is, for simplicity, taken
to be $\rho_{T}\left(0\right)=\left|l\right\rangle \left\langle l\right|\otimes\rho_{b}$,
i.e., the electron in the DQD localized in the left state $|l\rangle\langle l|$
and the bath in the thermal equilibrium state $\rho_{b}=e^{-\beta H_{b}}/\textrm{tr}_{b}e^{-\beta H_{b}}$.
The time-nonlocal master equation to second order in system-bath interaction
strength in the interaction picture reads \cite{Breuer02}
\begin{equation}
\frac{d}{dt}\tilde{\rho}_{s}\left(t\right)=-\int_{0}^{t}\textrm{tr}_{b}\left[\tilde{H}_{sb}\left(t\right),\left[\tilde{H}_{sb}\left(t'\right),\tilde{\rho}_{s}\left(t'\right)\otimes\rho_{b}\right]\right]dt',\label{weak master Eq interaction pic}
\end{equation}
where $\tilde{H}_{sb}\left(t\right)=\mathcal{G}_{s}\left(0,t\right)\sigma_{z}B\left(t\right)$,
$B\left(t\right)=\sum_{k}g_{k}\left(b_{k}^{\dagger}e^{i\omega_{k}t}+b_{k}e^{-i\omega_{k}t}\right)$.
The propagator superoperator $\mathcal{G}_{s}\left(t,t'\right)$ has
a general form of 
\begin{equation}
\mathcal{G}_{s}\left(t,t'\right)\equiv\textrm{T}_{+}\textrm{exp}\left[\intop_{t'}^{t}\mathscr{L}_{s}\left(t''\right)dt''\right],\label{eq:G}
\end{equation}
with $\textrm{T}_{+}$ denoting the time-ordering operator necessary
to allow an explicit time-dependent Hamiltonian \cite{Breuer02},
and the Liouville superoperator 
\begin{equation}
\mathscr{L}_{s}\left(t\right)A\equiv-i\left[H_{s}\left(t\right),A\right]\label{eq:L}
\end{equation}
is defined as the commutator between any operator $A$ and $H_{s}\left(t\right)$.

Performing the trace over the bath degrees of freedom and then going
back to the Schr\"odinger picture, one obtains from Eq.~(\ref{weak master Eq interaction pic})
\begin{equation}
\dot{\rho}_{s}\left(t\right)=\mathscr{L}_{s}\left(t\right)\rho_{s}\left(t\right)+\mathscr{L}_{z}\left[\mathcal{K}\left(t\right)+\textrm{H.c.}\right],\label{eq:wcME}
\end{equation}
where 
\begin{equation}
\mathcal{K}\left(t\right)=-i\int_{0}^{t}C\left(t-t'\right)\mathcal{G}_{s}(t,t')\sigma_{z}\rho_{s}\left(t'\right)dt',\label{eq:K(t)}
\end{equation}
$\textrm{H.c.}$ denotes the Hermitian conjugate of its previous term,
and $\mathscr{L}_{z}A=-i\left[\sigma_{z},A\right]$ for arbitrary
operator $A$, and the bath correlation function at temperature $k_{B}T=1/\beta$
is \cite{Carmichael,Breuer02} 
\begin{eqnarray}
C\left(\tau\right) & \equiv & \textrm{tr}_{b}\left[B\left(0\right)B\left(-\tau\right)\rho_{b}\right]\nonumber \\
 & = & \int_{0}^{\infty}d\omega J\left(\omega\right)\left[\cos\left(\omega\tau\right)\coth\left(\frac{\beta\omega}{2}\right)-i\sin\left(\omega\tau\right)\right],\nonumber \\
\label{C(tau)}
\end{eqnarray}
with the spectral density $J\left(\omega\right)=\sum_{k}\left|g_{k}\right|^{2}\delta\left(\omega-\omega_{k}\right)$.

To deal with Eqs.~(\ref{eq:wcME}) and (\ref{eq:K(t)}) without further
approximation, one can express the bath correlation function in terms
of a sum of exponentials as \cite{Optimal_control,Optimal_control2,initial_paper}:
\begin{equation}
C\left(\tau\right)=\sum_{m}\alpha_{m}e^{\gamma_{m}\tau},\label{eq:C_m}
\end{equation}
with complex numbers $\alpha_{m}$ and $\gamma_{m}$ that can be obtained
from numerical methods.

Substituting Eq.~(\ref{eq:C_m}) into Eq.~(\ref{eq:K(t)}), one
then obtains $\mathcal{K}\left(t\right)=\sum_{m}\mathcal{K}_{m}\left(t\right)$,
where $\mathcal{K}_{m}\left(t\right)=-i\int_{0}^{t}\alpha_{m}e^{\gamma_{m}\left(t-t'\right)}\mathcal{G}_{s}\left(t,t'\right)\sigma_{z}\rho_{s}\left(t'\right)dt'$.
By taking the time derivative of $\mathcal{K}_{m}\left(t\right)$
with the help of the property $\frac{\partial}{\partial t}\mathcal{G}_{s}\left(t,t'\right)=\mathscr{L}_{s}\left(t\right)\mathcal{G}_{s}\left(t,t'\right)$,
Eqs.~(\ref{eq:wcME}) and (\ref{eq:K(t)}) now become a set of linear
equations \cite{initial_paper}: 
\begin{eqnarray}
\dot{\rho}_{s}\left(t\right) & = & \mathscr{L}_{s}\left(t\right)\rho_{s}\left(t\right)+\mathscr{L}_{z}\sum_{m}\left[\mathcal{K}_{m}\left(t\right)+\textrm{H.c.}\right],\label{eq:wcME-sum}\\
\dot{\mathcal{K}}_{m}\left(t\right) & = & \left[\mathscr{L}_{s}\left(t\right)+\gamma_{m}\right]\mathcal{K}_{m}\left(t\right)-i\alpha_{m}\sigma_{z}\rho_{s}\left(t\right).\label{eq:dt_Km}
\end{eqnarray}
We have transformed the time-nonlocal master equation of Eqs.~(\ref{eq:wcME})
and (\ref{eq:K(t)}) into the time-local form of coupled Eqs.~(\ref{eq:wcME-sum})
and (\ref{eq:dt_Km}). These equations are valid even in a strong
driving field as the only approximation made in obtaining them is
the Born approximation in the weak system-bath coupling limit.

The bosonic bath considered in Refs. \cite{Stace_steady-state_PRL111_2013,step-like-experiment2014}
is a phonon bath and the spectral density for the piezoelectric phonon
coupling considered takes the form 
\begin{equation}
J\left(\omega\right)=\frac{P}{2}\frac{\omega\omega_{c}^{2}}{\omega^{2}+\omega_{c}^{2}}\left[1-\mathrm{sinc}\left(\frac{d\omega}{c_{s}}\right)\right],\label{J(w)}
\end{equation}
where $P$ is the piezoelectric electron-phonon coupling strength,
$\omega_{c}$ is the bath cutoff frequency, $d$ is the inter-dot
separation, and $c_{s}$ is the transverse speed of sound. The factor
$1-\mathrm{sinc}\left(d\omega/c_{s}\right)$ with $\mathrm{sinc}x=\sin x/x$
describes the oscillations on the frequency scale $c_{s}/d$ and leads
to deviations from the Lorentz-Drude spectral density.

Let us first discuss the behavior of the bath correlation function
with $J\left(\omega\right)$ given by Eq.~(\ref{J(w)}). The real
part of the bath correlation function Eq.~(\ref{C(tau)}) coming
from the $\mathrm{sinc}$ term of Eq.~(\ref{J(w)}) is convergent,
while the contribution coming from the Lorentz-Drude term (i.e., the
first term) of Eq.~(\ref{J(w)}) \cite{Breuer02}, 
\begin{equation}
\int_{0}^{\infty}d\omega\frac{P}{2}\omega\left[\frac{\omega_{c}^{2}}{\omega_{c}^{2}+\omega^{2}}\right]\cos\left(\omega\tau\right)\coth\left(\frac{\beta\omega}{2}\right)\label{eq:Re_C(t)_Lorentz-Drude}
\end{equation}
looks logarithmically divergent at $\tau=0$ for large frequencies.
However, if one does the time integral in Eq. (\ref{eq:K(t)}) first,
one will get additional $\omega^{-1}$ power and then the resultant
frequency integral will converge \cite{Lorentz_Drude_time_integral_converge}.
It is then reasonable to assume that the dynamics does not depend
appreciably on the very high-frequency bath modes. For our formulation,
we would like to evaluate the bath correlation function first and
then numerically fit it with multi-exponentials as in Eq.~(\ref{eq:C_m}).
We thus express the Lorentz-Drude spectral density as \cite{Lorentz-Drude_fitting,Lorentz-Drude_fitting_form}
\begin{equation}
\frac{P}{2}\frac{\omega\omega_{c}^{2}}{\omega^{2}+\omega_{c}^{2}}\sim\sum_{k}\frac{4p_{k}\Omega_{k}\omega}{\left(\omega^{2}-\Omega_{k}^{2}\right)^{2}+2\left(\omega^{2}+\Omega_{k}^{2}\right)\Gamma_{k}^{2}+\Gamma_{k}^{4}}\label{eq:J(w)_DL_fit}
\end{equation}
up to a sufficiently high frequency (e.g., up to $100\omega_{c}$),
where the fitting parameters $p_{k}$, $\Omega_{k}$, and $\Gamma_{k}$
are real numbers and can be obtained numerically. Then by substituting
Eq.~(\ref{eq:J(w)_DL_fit}) into Eq.~(\ref{J(w)}), a high-frequency
$\omega^{-4}$ power will converge the integral in Eq.~(\ref{C(tau)}).
Thus, expressing Eq.~(\ref{eq:C_m}) as a sum of exponentials becomes
achievable. We note that the poles of Eq.~(\ref{eq:J(w)_DL_fit})
correspond roughly to the poles of the response after a sequence of
approximations in the Laplace space in Ref.~\cite{Stace_steady-state_PRL111_2013}.
In the following section, we will introduce a full-polaron method
that is free of this divergence problem in its bath correlation functions
$C_{ij}(\tau)$. To verify that fitting Eq.~(\ref{eq:J(w)_DL_fit})
up to $100\omega_{c}$ is reasonable, we compare the case using directly
Eq.~(\ref{J(w)}) with that using Eq.~(\ref{eq:J(w)_DL_fit}) in
the full-polaron method, and find that both cases give the same dynamical
steady-state results presented in this paper.

\subsection{Polaron transformation}

\label{sec:Polaron_transformation}

To deal with the case of strong system-environment coupling, we first
make a polaron transformation to the model Hamiltonian Eq.~(\ref{H_T(t)})
by \cite{Criteria_for_small_polarion,VP_Master_Eq_DC_and_Imhomo}
\begin{equation}
H_{T}'\left(t\right)=e^{V}H_{T}\left(t\right)e^{-V},\label{eq:Var_polaron_transform}
\end{equation}
where $V=\frac{\Phi}{2}\sigma_{z}$, and 
\begin{equation}
\Phi\equiv2\sum_{k}\frac{g_{k}}{\omega_{k}}\left(b_{k}^{\dagger}-b_{k}\right).\label{eq:Phi}
\end{equation}
The transformed Hamiltonian can be written as 
\begin{eqnarray}
H_{T}'\left(t\right) & = & H_{s}'\left(t\right)+H_{sb}'\left(t\right)+H_{b},\label{H'_T(t)}\\
H_{s}'\left(t\right) & = & -\frac{\epsilon}{2}\sigma_{z}-\frac{\eta\Delta\left(t\right)}{2}\sigma_{x}-\sum_{k}\frac{g_{k}^{2}}{\omega_{k}}I,\label{H'_s(t)}\\
H_{sb}'\left(t\right) & = & -\frac{\Delta\left(t\right)}{2}\left(B_{x}\sigma_{x}+B_{y}\sigma_{y}\right),\label{H'_sb(t)}
\end{eqnarray}
where $I$ is an identity matrix, and 
\begin{eqnarray}
B_{x} & = & \left(\cosh\Phi-\eta\right),\label{eq:Bx}\\
B_{y} & = & i\sinh\Phi,\label{eq:By}
\end{eqnarray}
are the bath operators in the transformed frame. The parameter $\eta$
is defined as 
\begin{eqnarray}
\eta & = & \left\langle \cosh\Phi\right\rangle _{H_{b}}\nonumber \\
 & \equiv & \textrm{tr}_{b}\left(\rho_{b}\cosh\Phi\right)\nonumber \\
 & = & \exp\left[-2\sum_{k}\left(\frac{g_{k}}{\omega_{k}}\right)^{2}\coth\left(\frac{\beta\omega_{k}}{2}\right)\right].\label{eta(xi_k)}
\end{eqnarray}
We have subtracted $\eta$, with value $0\le\eta\le1$, from the bath
operator $B_{x}$ {[}see Eq.~(\ref{eq:Bx}{]} to make $\left\langle H_{sb}'\left(t\right)\right\rangle _{H_{b}}=0$,
and at the same time have added back a corresponding term $-\frac{\eta\Delta\left(t\right)}{2}\sigma_{x}$
to the system Hamiltonian, where $\Delta(t)$ is defined in Eq.~(\ref{eq:Delta(t)}).
This bath-renormalized tunneling term can describe the coherent dynamics
of the system at the phonon-dressed energy scale $\frac{\eta\Delta\left(t\right)}{2}$.
Reference \cite{Stace_steady-state_PRL111_2013} also has a scheme
to determine the renormalization of the Rabi frequency via a self-consistent
condition that the bath-renormalized system Hamiltonian in the Laplace
space should vanish in the interaction picture. The renormalization
of the coherent driving amplitude $\eta\Omega_{0}$ here is related
to the renormalization of the Rabi frequency in Ref.~\cite{Stace_steady-state_PRL111_2013}.
One can see this by expanding $\eta$ of Eq.~(\ref{eta(xi_k)}) to
second order in $g_{k}$ at zero temperature to obtain $\eta_{\textrm{2nd}}=1-2\int_{0}^{\infty}d\omega J\left(\omega\right)/\omega^{2}.$
Then $\eta_{\textrm{2nd}}\Omega_{0}$ is equal to the approximated
bath-induced renormalized Rabi frequency $\Omega_{\textrm{approx}}$
of Ref. \cite{Stace_steady-state_PRL111_2013}. This relation suggests
that their renormalization of the Rabi frequency contains the information
of the second-order system-bath coupling contribution while our full-polaron
one, i.e., $\frac{\eta\Delta\left(t\right)}{2}\sigma_{x}$, contains
not only the second-order but also higher-order contributions.

\subsection{Full-polaron master equation}

\label{sec:Full_polaron_master_equation_driven} Even though the original
system-bath interaction is strong, the new identified system-bath
interaction Hamiltonian $H_{sb}'(t)$ of Eq.~(\ref{H'_sb(t)}) that
depends on $\Delta\left(t\right)$ would be small and could be considered
as a perturbation term. Thus, once the parameter $\eta$ is determined
numerically, we then derive perturbatively in the transformed polaron
frame a quantum master equation to second order in $H_{sb}'(t)$ from
Eq.~(\ref{H'_T(t)}).

The model Hamiltonian, Eq.~(\ref{H_T(t)}), without the off-diagonal
term {[}i.e., $\Delta(t)=0${]} is a pure-dephasing spin-boson model,
and after the polaron transformation, the total Hamiltonian becomes
decoupled without the system-environment interaction in the transformed
frame. In this case, the reduced system dynamics is described by an
exact time-local (time-convolutionless) non-Markovian master equation.
So we will adopt for the $\Delta(t)\neq0$ case a time-local non-Markovian
master equation approach \cite{Breuer02,time-local_vs_non-loca_2nd_ME_initial}
to describe the time evolution of the reduced system density matrix,
\begin{equation}
\rho_{s}'\left(t\right)=\textrm{tr}_{b}\left[e^{V}\rho_{T}\left(t\right)e^{-V}\right]\label{rhos'(t)}
\end{equation}
in the polaron frame for our driven model. Because the initial total
density operator is chosen to be $\left|l\right\rangle \left\langle l\right|\otimes\rho_{b}$,
the polaron transformation displaces the initial bath state to $e^{V}\left(\left|l\right\rangle \left\langle l\right|\otimes\rho_{b}\right)e^{-V}=\left|l\right\rangle \left\langle l\right|\otimes\left(e^{\frac{\Phi}{2}}\rho_{b}e^{-\frac{\Phi}{2}}\right)$.
However, only the steady state is concerned in Refs. \cite{StaceDohertyBarrett2005,Stace_steady-state_PRL111_2013,step-like-experiment2014}
and the steady-state quantities are independent of the initial states,
i.e., does not depend on whether the bath state of the transformed
initial state is displaced or not \cite{Criteria_for_small_polarion,VP_Master_Eq_DC_and_Imhomo}.
As a result, we choose, for simplicity, the original undisplaced initial
state $\left|l\right\rangle \left\langle l\right|\otimes\rho_{b}$
as the initial state in the polaron frame. The full-polaron master
equation to second order in $\tilde{H}_{sb}'(t)$ in the interaction
picture with respect to $H_{0}'\left(t\right)=H_{s}'\left(t\right)+H_{b}$
reads \cite{driven_QD_RWA_small_polaron}
\begin{equation}
\frac{d}{dt}\tilde{\rho}_{s}'\left(t\right)=-\int_{0}^{t}\textrm{tr}_{b}\left[\tilde{H}_{sb}'\left(t\right),\left[\tilde{H}_{sb}'\left(t'\right),\tilde{\rho}_{s}'\left(t\right)\otimes\rho_{b}\right]\right]dt',\label{FP master Eq interaction pic}
\end{equation}
where $\tilde{H}_{sb}'\left(t\right)=\mathcal{G}_{0}'\left(0,t\right)H_{sb}'\left(t\right)$.
The propagator superoperator is defined as 
\begin{equation}
\mathcal{G}_{j}'\left(t,t'\right)\equiv\textrm{T}_{+}\exp\left[\intop_{t'}^{t}\mathscr{L}_{j}'\left(t''\right)dt''\right]\label{eq:G-1}
\end{equation}
and the Liouville superoperator 
\begin{equation}
\mathscr{L}_{j}'\left(t\right)A\equiv-i\left[H_{j}'\left(t\right),A\right]\label{eq:L-1}
\end{equation}
is defined as a commutator between any operator $A$ and $H_{j}'\left(t\right)$
with $j=0$ for the present case. Later we will introduce $\mathscr{L}_{s}'\left(t\right)$
and its corresponding $\mathcal{G}_{s}'\left(t,t'\right)$ as defined
in Eqs.~(\ref{eq:L-1}) and (\ref{eq:G-1}) for the system alone
with the replacement of Hamiltonian $H_{j}'\left(t\right)\to H_{s}'\left(t\right)$.

Defining $\sigma_{1}=\sigma_{x}$, $\sigma_{2}=\sigma_{y}$, $B_{1}=B_{x}$,
and $B_{2}=B_{y}$, performing the trace over the bath degrees of
freedom and then going back to the Schr\"odinger picture, one can
write a concise expression for the second-order time-local master
equation from Eq.~(\ref{FP master Eq interaction pic}) as 
\begin{equation}
\dot{\rho}_{s}'\left(t\right)=\mathscr{L}_{s}'\left(t\right)\rho_{s}'\left(t\right)-\sum_{i,j}\left[\frac{\Delta\left(t\right)}{2}\mathscr{L}_{i}D_{ij}\left(t\right)\rho'_{s}\left(t\right)+\textrm{H.c.}\right],\label{time-dep FP master eq}
\end{equation}
where 
\begin{equation}
D_{ij}\left(t\right)=i\int_{0}^{t}C_{ij}\left(\tau\right)\frac{\Delta\left(t'\right)}{2}\mathcal{G}_{s}'\left(t,t'\right)\sigma_{j}dt',\label{D_ij(t)}
\end{equation}
with $i,j$ running from $1$ to $2$, and $\mathscr{L}_{i}A=-i\left[\sigma_{i},A\right]$
for any operator $A$. The explicit expressions for the bath correlation
functions $C_{ij}\left(\tau\right)=\left\langle B_{i}B_{j}\left(-\tau\right)\right\rangle _{H_{b}}=\textrm{tr}_{b}\left[B_{i}e^{-iH_{b}\tau}B_{j}e^{iH_{b}\tau}\rho_{b}\right]$
are \cite{driven_QD_RWA_small_polaron} 
\begin{eqnarray}
C_{11}\left(\tau\right) & = & \eta^{2}\left\{ \cosh\left[r\left(\tau\right)\right]-1\right\} ,\label{eq:C11}\\
C_{22}\left(\tau\right) & = & -\eta^{2}\sinh\left[r\left(\tau\right)\right],\label{eq:C22}
\end{eqnarray}
where 
\begin{eqnarray}
r\left(\tau\right) & = & -4\int_{0}^{\infty}d\omega\frac{J\left(\omega\right)}{\omega^{2}}\nonumber \\
 &  & \times\left[\cos\left(\omega\tau\right)\coth\left(\frac{\beta\omega}{2}\right)-i\sin\left(\omega\tau\right)\right].\label{r(tau)}
\end{eqnarray}
The other bath cross correlation functions vanish, i.e., $C_{12}\left(\tau\right)=C_{21}\left(\tau\right)=0$.
Given Eq.~(\ref{J(w)}), both Eqs.~(\ref{eq:C11}) and (\ref{eq:C22})
converge at $\tau\geq0$ for both $\omega\rightarrow\infty$ (due
to $\omega^{-2}$) and $\omega\rightarrow0$ (due to $\left[1-\mathrm{sinc}\left(d\omega\right)\right]/\omega^{2}$)
in the integral of Eq.~(\ref{r(tau)}). So there is no divergence
problem to express, similar to the weak-system-bath-coupling case,
each of the bath correlation functions as a sum of exponentials as:
\begin{equation}
C_{ij}\left(\tau\right)=\sum_{m}\alpha_{ij,m}e^{\gamma_{ij,m}\tau},\label{C_m,ij VP}
\end{equation}
with complex numbers $\alpha_{ij,m}$ and $\gamma_{ij,m}$ obtained
by numerical methods. This enables us to verify the validity of the
expression of Eq.~(\ref{eq:J(w)_DL_fit}), and we find that using
Eq.~(\ref{eq:J(w)_DL_fit}) to replace the Lorentz-Drude term gives
the same population dynamics and the same time-averaged steady-state
results presented here as those obtained directly using Eq.~(\ref{J(w)}).

Inserting Eq.~(\ref{C_m,ij VP}) into Eq.~(\ref{D_ij(t)}), one
obtains $D_{ij}\left(t\right)=\sum_{m}D_{ij,m}\left(t\right)$, where
\begin{equation}
D_{ij,m}\left(t\right)=i\int_{0}^{t}\alpha_{ij,m}e^{\gamma_{ij,m}\tau}\frac{\Delta\left(t'\right)}{2}\mathcal{G}_{s}'\left(t,t'\right)\sigma_{j}dt'.\label{D_ij,m(t)}
\end{equation}
After taking the time derivative of Eq.~(\ref{D_ij,m(t)}), the resultant
equation together with Eq.~(\ref{time-dep FP master eq}) form a
set of differential equations: 
\begin{eqnarray}
\dot{\rho}_{s}'\left(t\right) & = & \mathscr{L}_{s}'\left(t\right)\rho_{s}'\left(t\right)\nonumber \\
 &  & -\sum_{i,j,m}\left\{ \frac{\Delta\left(t\right)}{2}\mathscr{L}_{i}D_{ij,m}\left(t\right)\rho'_{s}\left(t\right)+\textrm{H.c.}\right\} ,\label{time-dep FP master eq sum_m}
\end{eqnarray}
\begin{equation}
\dot{D}_{ij,m}\left(t\right)=\left[\mathscr{L}_{s}'\left(t\right)+\gamma_{ij,m}\right]D_{ij,m}\left(t\right)+i\alpha_{ij,m}\frac{\Delta\left(t\right)}{2}\sigma_{j}.\label{dot_D_ij,m(t)}
\end{equation}
Equation (\ref{time-dep FP master eq}) or the set of Eqs. (\ref{time-dep FP master eq sum_m})
and (\ref{dot_D_ij,m(t)}) is the full-polaron master equation that
will be used for dealing with time-dependent driving field problems,
without making both the rotating-wave approximation and the Markovian
approximation.

\section{Numerical results}

\label{sec:Results}

\begin{figure*}
\includegraphics[width=1\textwidth]{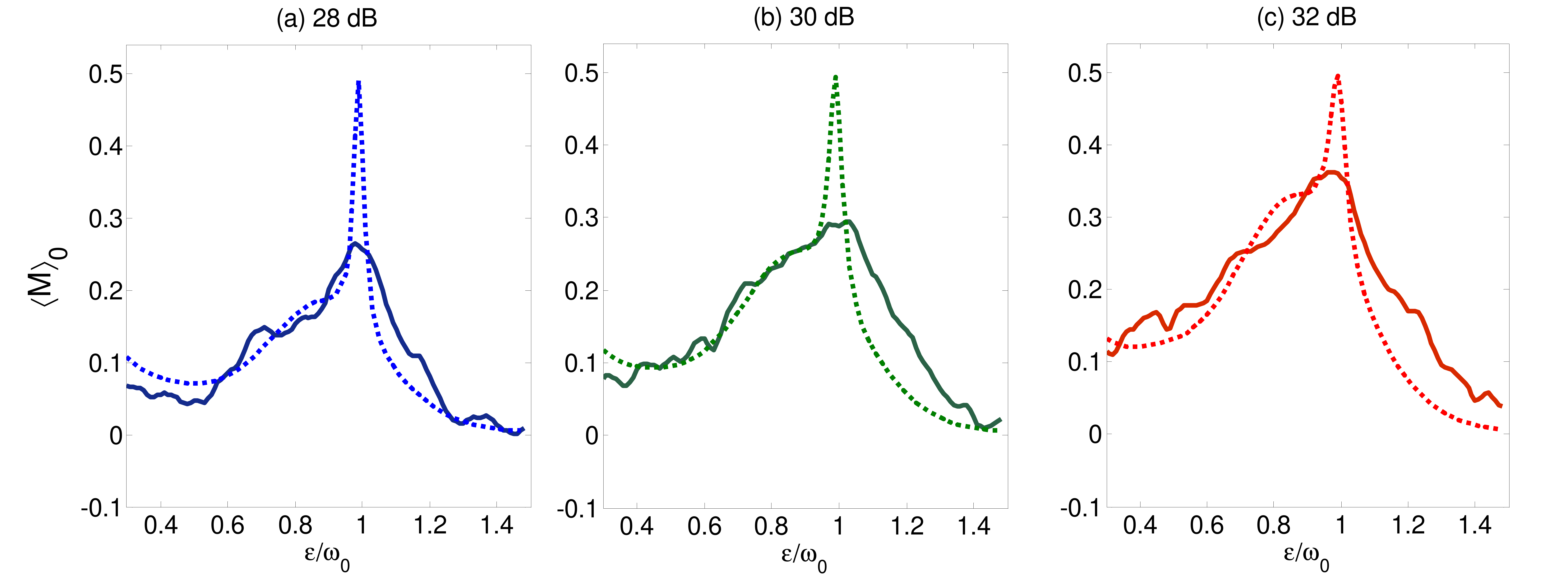}\protect\caption{Comparison of time-averaged steady-state population of the right dot
$\left\langle M\right\rangle _{0}=\left\langle \left|r\right\rangle \left\langle r\right|\right\rangle _{0}$
as a function of bias $\epsilon/\omega_{0}$ normalized by the driving
microwave angular frequency $\omega_{0}=2\pi\times32\textrm{ GHz}$
in the driven DQD system for different driving strengths of $\Omega_{0}=0.034\times10^{n}\omega_{0}$
where $n=\left\{ 0,0.1,0.2\right\} $ correspond to (a) 28 dB, (b)
30 dB, and (c) 32 dB, respectively. The experimental data taken from
Fig.~4 of Ref.~\cite{step-like-experiment2014} are replotted in
solid lines, and the results obtained from the full-polaron master
equation method are in dotted lines. The value of interdot separation
$d/c_{s}=16/\omega_{0}$, is the fitting value obtained from the full-polaron
master equation method. Other parameters used are $P=0.09$, $\Delta=0.15\omega_{0}$,
$\omega_{c}=2\omega_{0}$, and $k_{B}T=0.12\omega_{0}$.}
\label{fig:M_0_d_16_fit} 
\end{figure*}

The key quantities we will calculate and compare are the time-averaged
steady-state values of the right dot population $\left\langle M\right\rangle _{0}=\left\langle \left|r\right\rangle \left\langle r\right|\right\rangle _{0}=\left\langle \left(1-\sigma_{z}\right)/2\right\rangle _{0}$
\cite{Stace_steady-state_PRL111_2013}. The electron charge state
or population of the DQD system can be read out by a rf quantum point
contact (rf-QPC) in the experiment \cite{step-like-experiment2014}.
To check whether the interaction strength between the DQD and the
phonon bath is in the weak-coupling regime, we use the weak-coupling
master equation Eq. (\ref{eq:wcME-sum}) to obtain the time evolutions
of the population $\textrm{tr}_{T}\left[\left|r\right\rangle \left\langle r\right|\rho_{T}\left(t\right)\right]=\textrm{tr}_{s}\left[\left|r\right\rangle \left\langle r\right|\rho_{s}\left(t\right)\right]$,
and the full-polaron master equation Eq. (\ref{time-dep FP master eq sum_m})
to obtain 
\begin{eqnarray}
\textrm{tr}_{T}\left[\left|r\right\rangle \left\langle r\right|\rho_{T}\left(t\right)\right] & = & \textrm{tr}_{T}\left[\left|r\right\rangle \left\langle r\right|e^{V}\rho_{T}\left(t\right)e^{-V}\right]\nonumber \\
 & = & \textrm{tr}_{s}\left[\left|r\right\rangle \left\langle r\right|\rho_{s}'\left(t\right)\right].
\end{eqnarray}
We wait until the time evolution of the right dot population has reached
the steady state, i.e., a steady sinusoidal oscillation, and then
take the time average value in the steady state to obtain $\left\langle M\right\rangle _{0}$.
In this paper, we refer to the approach using Eq.~(\ref{eq:wcME-sum})
to calculate the results as the weak-coupling treatment and the approach
using Eq.~(\ref{time-dep FP master eq sum_m}) as the full-polaron
method. Using the original parameter set adopted in Fig.~4 of Ref.~\cite{step-like-experiment2014},
we found that the results of $\left\langle M\right\rangle _{0}$ obtained
by our full-polaron method and the weak-coupling treatment cannot
fit well the corresponding experimental data. Moreover, the weak-coupling
treatment even gives unphysical negative populations. A more detailed
comparison and description about this can be found in Appendix \ref{sec:appendixA_d_20_fit}.
If we decrease the system-bath coupling strength from $P=0.09$ to
a small enough value, e.g., $P=0.008$, the results of the full-polaron
method and the weak-coupling treatment approach each other without
any negative populations but still deviate from the experimental data
{[}see Figs.~\ref{fig:steady_state_population 2}(d)--\ref{fig:steady_state_population 2}(f)
in Appendix \ref{sec:appendixA_d_20_fit} where a simple estimation
of the validity criterion for the weak-coupling treatment is also
given{]}. We discuss furthermore in Appendix \ref{sec:appendix_positivity}
the loss of positivity (negative population) problem in second-order
master equations and its indication to the breakdown of the weak-coupling
assumption. These results suggest that the original parameter set
is not in the weak-coupling regime and may require some adjustment.

We notice that except tunneling rate and temperature obtained independently
from the experimental data \cite{step-like-experiment2014}, the other
parameters are estimated microscopically in Ref. \cite{step-like-experiment2014}
and may be adjusted slightly to obtain a better fitting to the experimental
data. We find that with the other original parameters remaining unchanged,
by adjusting the interdot separation slightly from $d/c_{s}=20/\omega_{0}$
to $d/c_{s}=14.8/\omega_{0}\sim16/\omega_{0}$ (with the free parameter
$\Omega_{0}$ adjusted correspondingly to $\Omega_{0}=0.038\omega_{0}\sim0.034\omega_{0}$),
a better fit can be obtained. We show for the case of $d/c_{s}=16/\omega_{0}$
in Figs.~\ref{fig:M_0_d_16_fit}(a)--\ref{fig:M_0_d_16_fit}(c) the
results of the time-averaged steady-state values of the right dot
population $\left\langle M\right\rangle _{0}$ as a function of bias
$\epsilon/\omega_{0}$ for driving field strengths $\Omega_{0}=0.034\times10^{n}\omega_{0}$,
where $n=\left\{ 0,0.1,0.2\right\} $ corresponds to 28 dB, 30 dB,
and 32 dB, respectively, in Ref.~\cite{step-like-experiment2014}.
The solid lines in Fig.~\ref{fig:M_0_d_16_fit} are the experimental
data in Fig.~4 of Ref.~\cite{step-like-experiment2014} and the
dotted lines represent the full-polaron results. Our full-polaron
method can fit well the steplike shoulders on the blue-detuned side
of the asymmetric resonance profile. Similar to the fitting result
at 32 dB by the theoretical method \cite{Stace_steady-state_PRL111_2013}
presented in Fig.~4 of Ref.~\cite{step-like-experiment2014}, the
fitting result by the full-polaron method (dotted line) in Fig.~\ref{fig:M_0_d_16_fit}(c)
shows a little higher shoulder than the experimental data (solid line).
Taking $c_{s}=3000\textrm{ ms}^{-1}$ for $d/c_{s}=16/\omega_{0}$,
we find $d\sim240\textrm{ nm}$. This value of $d$ is within the
possible value of the distance between the localized states of the
DQD confined by the surface gates in Ref.~\cite{step-like-experiment2014}.
We also estimate the interdot separation from the multi-phonon excitation
process instead of the single-phonon excitation process in the weak-coupling
case and obtain a value consistent with that obtained from our full-polaron
result of $d/c_{s}=16/\omega_{0}$ \cite{d_multi_phonon}.

Our presented results of the time-averaged steady-state population
all have sharp resonance peaks at the value about 0.5. However, as
described in Ref. \cite{step-like-experiment2014}, the existence
of charge noise, when averaged by the rf-QPC, rounds out or smears
out the sharper features of the resonance peaks in the experimental
data if the resolution around the resonance peaks is not high enough.
Therefore, the focus will be on the overall population behavior rather
than on the resonance peak values. The line shapes of the resonance
peaks exhibit strong phonon-induced and driving-induced asymmetry.
The enhanced population in the blue-detuned side of the peaks, where
the microwave photon energy exceeds the qubit splitting, is the consequence
of photon absorption from the driving field accompanied by a Raman
phonon emission, leading to a higher rate of excitation than the relaxation
rate \cite{StaceDohertyBarrett2005}.

\section{Discussions}

\label{sec:Discussion} We briefly discuss and compare the theoretical
approach used in Refs.~\cite{Stace_steady-state_PRL111_2013,step-like-experiment2014}
with our weak-coupling and full-polaron master equation approaches
here. The theoretical method in Refs.~\cite{step-like-experiment2014,Stace_steady-state_PRL111_2013}
involves a Laplace transformation to a second-order time-nonlocal
non-Markovian master equation and an energy renormalization scheme
with perturbative contributions from the interaction with the bath.
So the critical difference between their method and our weak-coupling
treatment is their additional renormalization scheme. By comparing
their theoretical results in Fig.~4 of Ref.~\cite{step-like-experiment2014}
with our weak-coupling results, the obvious effect of their renormalization
scheme is the correction of the negative populations when the parameter
set ($P=0.09$) is beyond the weak-coupling regime.  References \cite{Stace_steady-state_PRL111_2013,step-like-experiment2014}
show also that the steplike shoulders come from the spectral density.
By expressing $J\left(\omega\right)$ as Eq.~(\ref{eq:J(w)_DL_fit}),
our weak-coupling treatment can also have the steplike shoulder behaviors
on all the results {[}see Fig. \ref{fig:steady_state_population 2}
in Appendix \ref{sec:appendixA_d_20_fit}{]}. In Fig.~\ref{fig:M_0_d_16_fit},
to fit the steplike shoulders we have decreased the interdot separation
slightly from $d/c_{s}=20/\omega_{0}$ to $d/c_{s}=16/\omega_{0}$
for the full-polaron method. We have also checked that the weak-coupling
treatment breaks down with negative populations for these parameters
because $P=0.09$ is not small enough.

We discuss next how the perturbative renormalization scheme employed
in Refs.~\cite{Stace_steady-state_PRL111_2013,step-like-experiment2014}
can help mitigate the positivity violation problem than the traditional
perturbative approach. The positivity violation (negative population
values) problem for traditional perturbative second-order master equations
is discussed in Appendix \ref{sec:appendix_positivity}. The perturbative
renormalization scheme in Refs.~\cite{Stace_steady-state_PRL111_2013,step-like-experiment2014}
is slightly different from the traditional perturbative approach and
is achieved by going to a special basis of the interaction picture
via a suitable choice of the dressing Hamiltonian $H_{D}$ to cancel
the bath-induced dispersive shifts (i.e., imaginary parts of the second-order
perturbation kernels arising from the bath) so the renormalized system
Hamiltonian vanishes in this interaction picture. This determines
the dressing system Hamiltonian $H_{D}$ with two second-order renormalized
energies: the diagonal detuning arising from the bath-induced Lamb
shift and the off-diagonal Rabi frequency also arising from the bath-induced
contribution, which reduces the transition dipole moments. This in
turn gives a system energy in the dressed basis closer to the real
open system energy than the bare system eigenenergy in the interaction
picture of the free Hamiltonian. In summary, in the perturbative scheme
(in the system-bath coupling) employed in Refs.~\cite{Stace_steady-state_PRL111_2013,step-like-experiment2014},
the steady-state solutions with second-order corrections to the diagonal
elements of the density matrix operator come from two sources. One
is from the bath-induced renormalized system Hamiltonian which constitutes
the zeroth-order nonperturbative Hamiltonian. This non-perturbative
correction is not accessible in the traditional second-order perturbative
master equation that works with a bare zeroth-order system Hamiltonian.
The other one is from the remaining second-order perturbation kernels
(non-Markovian and non-Lindblad form) that are not canceled in the
interaction picture determined by the dressing Hamiltonian. It is
the perturbative correction that may induce the second-order positivity
violation. But due to the transformation to the basis with respect
to the dressing Hamiltonian, the parameter values of their second-order
perturbation scheme are altered, resulting in smaller magnitudes of
the perturbative kernels or a better perturbation scheme than the
traditional weak-coupling perturbation method. Thus smaller fourth-order
contributions required for a full second-order solution in the perturbative
scheme with renormalized energies employed in Refs.~\cite{Stace_steady-state_PRL111_2013,step-like-experiment2014}
are expected. As a result, it helps mitigate the positivity violation
problem or can tolerate a larger parameter regime than the validity
regime of the traditional master equation approach. In other words,
the renormalization scheme helps lessen the problem of negative right-dot
populations for the parameter set in Ref.~\cite{step-like-experiment2014},
which is considered beyond the weak-coupling regime. If the system-environment
interaction is increased further, this approach will eventually also
give negative populations although a less negative value than that
by the traditional master equation approach is anticipated.

In contrast, the full-polaron master equation is designed to deal
with the strong system-environment coupling case and thus can sustain
validity over a much wider parameter regime than the perturbative
schemes. The polaron transformation, Eqs.~(\ref{eq:Var_polaron_transform})
and (\ref{eq:Phi}), has a nonperturbative nature in the system-environment
coupling. After the polaron transformation, pure-electronic modes
are renormalized to polaronic modes with larger effective mass, reducing
the coherent tunneling or driving term. The transformed system-bath
interaction Hamiltonian that depends on the coherent tunneling or
driving term becomes small in the polaron frame and can be treated
with regular perturbative master equation approach effectively. Thus
the full-polaron master equation can remain valid and go beyond the
positivity violation problem for a wider parameter region of the system-environment
coupling strength.

Indeed, one can see that there are tiny negative values in the time-averaged
steady-state population beside the extra small peaks on the right-hand
sides of the dashed lines of the theoretical curves near $(\epsilon/\omega_{0})=1.4$
in Figs.~4(a)--4(c) of Ref.~\cite{step-like-experiment2014}, but
they are not present in our full-polaron master equation results shown
in Fig.~\ref{fig:M_0_d_16_fit}. This demonstrates that although
their renormalization scheme can tolerate a larger parameter regime
than our weak-coupling master equation approach {[}see Figs.~\ref{fig:steady_state_population 2}(a)--\ref{fig:steady_state_population 2}(c)
in Appendix \ref{sec:appendixA_d_20_fit}{]}, it still leaves a tiny
violation in positivity in that small parameter region. In other words,
while the renormalization terms in Refs.~\cite{Stace_steady-state_PRL111_2013,step-like-experiment2014}
contain the contributions from the second-order system-bath coupling,
which help lessen the negative population problem, our full-polaron
method also contains higher orders as in Eq.~(\ref{eta(xi_k)}) and
can also capture the effect of multi-bath-quanta processes when the
system-bath coupling is not weak, thus capable of going beyond the
positivity violation problem for a wider parameter region.

\section{Conclusion}

\label{sec:Conclusion}

We have presented a full-polaron master equation and a weak-coupling
master equation to describe the steady-state time-averaged electron
population of a driven DQD system interacting with a phonon bath and
compare the obtained results with those from a recent experiment and
its corresponding theoretical method. We find that the original parameter
set used in their experiment and theoretical method is beyond the
weak-coupling parameter regime. By using our full-polaron method with
a slight change of a single parameter of interdot distance from $d=20c_{s}/\omega_{0}$
to $d=16c_{s}/\omega_{0}$, the experimental results of steplike shoulder
behaviors can be fitted rather well. Our full-polaron equation approach
does not require the renormalization scheme employed in their weak-coupling
theory \cite{Stace_steady-state_PRL111_2013}, and can still describe
the driving-induced phonon-enhanced phenomena in the experiment. The
full-polaron and weak-coupling master equations presented here are
efficient tools and can be used to describe, over a quite wide range
of parameters in the parameter space, the steady-state behaviors of
a driven open quantum system.
\begin{acknowledgments}
H.S.G. acknowledges support from the the Ministry of Science and Technology
of Taiwan under Grants No.~MOST 106-2112-M-002-013-MY3, No.~MOST
108-2627-E-002-001 and No.~MOST 108-2622-8-002-016, from the National
Taiwan University under Grants No.~NTU-CC-108L893202 and No.~NTU-CC-109L892002,
and from the thematic group program of the National Center for Theoretical
Sciences, Taiwan. T.M.S. was supported by the Australian Research
Council Centres of Excellence for Engineered Quantum Systems (EQUS,
No.~CE170100009). 
\end{acknowledgments}

\appendix

\section{Validity of the parameter set of the driven DQD system used in Ref.~\cite{step-like-experiment2014}}

\label{sec:appendixA_d_20_fit}

\begin{figure*}
\includegraphics[width=1\textwidth]{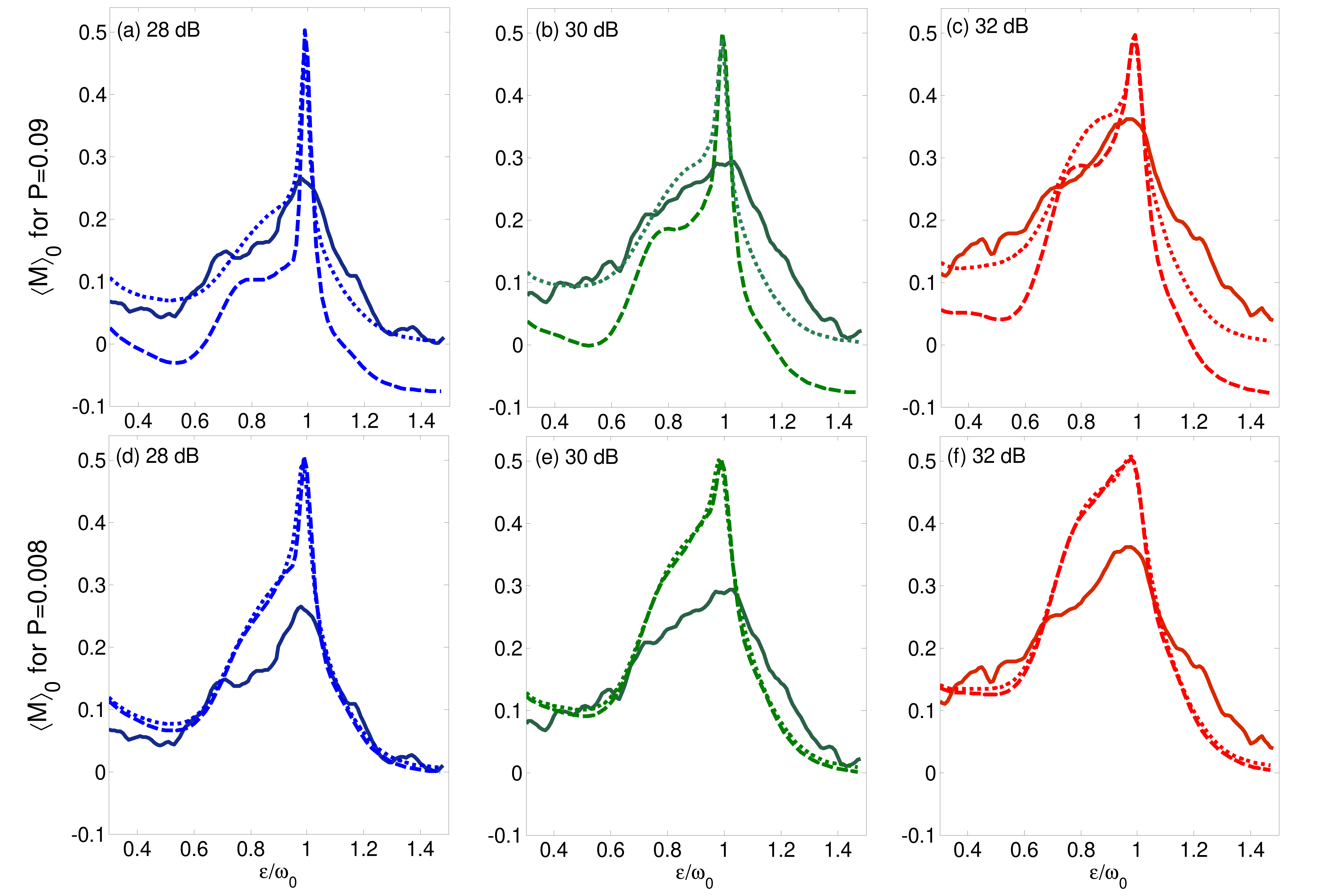}\protect\caption{Comparison of time-averaged steady-state population of the right dot
$\left\langle M\right\rangle _{0}=\left\langle \left|r\right\rangle \left\langle r\right|\right\rangle _{0}$
as a function of bias $\epsilon/\omega_{0}$ normalized by the driving
microwave angular frequency $\omega_{0}=2\pi\times32\textrm{ GHz}$
in the driven DQD system for different driving strengths of $\Omega_{0}=0.034\times10^{n}\omega_{0}$,
where $n=\left\{ 0,0.1,0.2\right\} $ correspond to 28 dB {[}left
panel, (a) and (d){]}, 30 dB {[}central panel, (b) and (e){]} and
32 dB {[}right panel, (c) and (f){]}, respectively. The experimental
data taken from Fig.~4 of Ref.~\cite{step-like-experiment2014}
are replotted in solid lines, and the results obtained from the full-polaron
and the weak-coupling master equations are in dotted lines and dashed
lines, respectively. The value of interdot separation $d/c_{s}=20/\omega_{0}$
is the original parameter value in Ref.~\cite{step-like-experiment2014}.
The original value of $P=0.09$ given in Ref.~\cite{step-like-experiment2014}
is used in (a)--(c). If $P=0.008$ is an order of magnitude smaller,
the results of the full-polaron and the weak-coupling methods as shown
in (d)--(f) are in good agreement with each other. Other parameters
used are $\Delta=0.15\omega_{0}$, $\omega_{c}=2\omega_{0}$, and
$k_{B}T=0.12\omega_{0}$.}
\label{fig:steady_state_population 2} 
\end{figure*}

In this Appendix, we discuss how the original parameter set used in
the theoretical method \cite{Stace_steady-state_PRL111_2013} to explain
the experiment of the driven DQD system in Ref.~\cite{step-like-experiment2014}
is not in the weak-coupling regime and may require some adjustment.
Figure \ref{fig:steady_state_population 2} shows the time-averaged
steady-state values of the right dot population $\left\langle M\right\rangle _{0}$
as a function of bias $\epsilon/\omega_{0}$ for driving field strengths
$\Omega_{0}=0.034\times10^{n}\omega_{0}$, where $n=\left\{ 0,0.1,0.2\right\} $
corresponds to 28 dB {[}left panel, (a) and (d){]}, 30 dB {[}central
panel, (b) and (e){]} and 32 dB {[}right panel, (c) and (f){]}, respectively,
in Ref.~\cite{step-like-experiment2014}. The solid lines (blue,
green, and red in the top panel are for the convenience of comparison
the same as those in the bottom panel and in Fig.~\ref{fig:M_0_d_16_fit})
in Fig.~\ref{fig:steady_state_population 2} are the experimental
data in Fig.~4 of Ref.~\cite{step-like-experiment2014}.

By using the original parameter set adopted in Fig.~4 of Ref.~\cite{step-like-experiment2014}
{[}$\Omega_{0}$ is a free parameter but the ratio of one value to
another on a logarithmic scale in decibel (dB) unit is fixed{]}, Figs.~\ref{fig:steady_state_population 2}(a)--(c)
show the results of our full-polaron method (dotted lines), the weak-coupling
treatment (dashed lines), and the experimental data (solid lines).
One can see that they deviate from each other and the weak-coupling
treatment even gives unphysical negative populations. Larger deviations
between the results of the weak-coupling treatment and the full-polaron
method occur in the off-resonance regimes rather than near the resonance
peaks. One might be tempted to think that the weak-coupling treatment
is close to the full-polaron method near the resonance peaks. In fact,
the detailed dynamics of the weak-coupling treatment deviates from
that of the full-polaron method. In other words, because only the
time-averaged steady-state populations are compared there, the weak-coupling
treatment gives a close result to that of the full-polaron method.
But it does not really mean that the weak-coupling treatment is valid
near resonance for this set of parameters. In short, these results
suggest that the original parameter set is not for processing in the
weak-coupling regime because the weak-coupling treatment breaks down.

Actually, the weak-coupling treatment is expected to be valid when
the criterion $\max(\Omega_{0},|\delta|)\gg\Gamma$ is satisfied,
where the detuning $\left|\delta\right|=\left|W-\omega_{0}\right|$
and $W=\sqrt{\epsilon^{2}+\Delta^{2}}$. We have obtained $\Omega_{0}\approx0.034\omega_{0}$
in Figs.~\ref{fig:M_0_d_16_fit} and \ref{fig:steady_state_population 2},
so in the regime of appreciable off-resonance, we have $|\delta|>\Omega_{0}$.
As a result, the validity criterion of the weak-coupling treatment
should require the decay rate $\Gamma\ll0.034\omega_{0}$. The decay
rate $\Gamma$ can be estimated by the expression of the coefficient
of the decay term in the master equation as $\Gamma/2\sim\int_{0}^{\omega_{c}^{-1}}C\left(0\right)dt'$.
In estimating the bath correlation function $\left|C\left(0\right)\right|$,
the $\mathrm{sinc}\left(d\omega/c_{s}\right)$ term in the bath spectral
density $J\left(\omega\right)$ with parameters $d/c_{s}=14.8/\omega_{0}\sim20/\omega_{0}$
has small contributions and can be neglected. So in the low- or zero-temperature
limit, we have $C\left(0\right)\sim\int_{0}^{\omega_{c}}\left(P\omega/2\right)d\omega=P\omega_{c}^{2}/4$,
and this leads to $\Gamma\sim P\omega_{0}$. To satisfy the validity
criterion of the weak-coupling treatment of $\Gamma\ll0.034\omega_{0}$,
we choose as an example $P=0.008$, one order of magnitude smaller
than $P=0.09$ in the original parameter set, and show the results
in Figs.~\ref{fig:steady_state_population 2}(d)--\ref{fig:steady_state_population 2}(f).
One can see that the results of the weak-coupling treatment approach
to that of the full-polaron method without any breakdown.

However, since no result in Figs.~\ref{fig:steady_state_population 2}(a)--\ref{fig:steady_state_population 2}(c)
can fit well its corresponding experimental data (solid line), we
conclude that the original parameter set is not in the weak-coupling
regime and may require some adjustment.

\section{Positivity violations in second-order master equations}

\label{sec:appendix_positivity}

We briefly discuss the problem of negative population results of the
weak-coupling master equation in this Appendix. 

A density matrix $\rho$ should be positive semi-definite, i.e., $\langle x|\rho|x\rangle\geq0$
for all states $|x\rangle$. The solution of the reduced system density
matrix $\rho_{s}(t)$ in the master-equation approach for an open
quantum system can be guaranteed to be positive semidefinite at all
times if the master equation is exact or is of Lindblad form. It has
been shown that perturbative second-order (in system-environment coupling
strength) time-local \cite{positivity_2nd_MasterEquation_Fleming2011,positivity_2nd_MasterEquation_Fleming2012,FlemingHu2012,Modified_redfield_2012,positivity_is_weak-coupling_2020}
or time-nonlocal \cite{TC_master_eq_positivity_BarnettStenholm2001,Breuer02,TC_master_eq_positivity_Budini2004,TC_master_eq_positivity_Maniscalco2005,TC_master_eq_positivity_BreuerVacchini2009,TC_master_eq_positivity_VacchiniBreuer2010,positivity_2nd_MasterEquation_Fleming2011,FlemingHu2012}
master equations may not guarantee yielding a dynamical map with exact
complete positivity. In other words, these second-order master equations
may not ensure completely positive evolution and may give unphysical
negative eigenvalues of the density-matrix operator after some time
if the parameters are beyond their range of validity.

References \cite{positivity_2nd_MasterEquation_Fleming2011,positivity_2nd_MasterEquation_Fleming2012,Modified_redfield_2012}
have shown that the long-time dynamics of order-2n accuracy of a perturbative
density-matrix operator requires an order-$(2n+2)$ master equation.
In other words, a perturbative master equation to second order in
the system-bath coupling strength yields a full-time solution of the
density-matrix operator with accuracy of zeroth order. This can lead
to second-order violations of positivity in long-time (steady-state)
regimes, especially at low temperatures, as the diagonal elements
of the reduced density matrix in the energy basis of the free Hamiltonian
are not perturbed to the correct second-order values \cite{positivity_2nd_MasterEquation_Fleming2011,positivity_2nd_MasterEquation_Fleming2012,FlemingHu2012,Modified_redfield_2012}.
Reference \cite{positivity_is_weak-coupling_2020} has shown that
the positivity violations in the Redfield master equation with time-dependent
coefficients (similar to our weak-coupling master equation) occur
only in a parameter regime where the perturbative Redfield master
equation becomes significantly invalid, i.e., in a parameter regime
of larger system-bath coupling strength or larger bath correlation
time. This implies that the loss of positivity should in fact be welcomed
as an important feature: It indicates the breakdown of the weak-coupling
assumption \cite{positivity_is_weak-coupling_2020}.

To correct the positivity problem, one should require a full second-order
solution for the diagonal elements (populations) to keep the density-matrix
operator positive semi-definite. This can be achieved consistently
to second order from a perturbative process through the help of the
fourth-order master equation, or a perturbative expansion of the exact
solution to second order, or from a nonperturbative process described
by a Lindblad master equation.

On the other hand, the perturbative renormalization scheme employed
in Refs.~\cite{Stace_steady-state_PRL111_2013,step-like-experiment2014}
and our full-polaron method can help mitigate the positivity violation
problem as compared to the traditional perturbative approach (see
Sec.~\ref{sec:Discussion} in the main text for a brief discussion).

\bibliographystyle{apsrev4-1}
\bibliography{dynamical_steady_state_Full_polaron_reference}

\end{document}